\documentclass[prb,twocolumn,longbibliography,floatfix]{revtex4-1}
\usepackage[T1]{fontenc}
\usepackage[utf8]{inputenc}
\usepackage[a4paper]{geometry}
\usepackage{hyperref}
\usepackage{graphicx}
\usepackage{xcolor}
\usepackage{multirow}

\usepackage[caption=false]{subfig}

\usepackage[charter]{mathdesign}

\makeatletter
\@ifundefined{date}{}{\date{}}
\makeatother

\usepackage{amsmath}
\usepackage{mathtools}
\usepackage{physics}
\usepackage{bbm}
\usepackage{pifont} 

\usepackage{soul} 

\newcommand{\ncmd}{\newcommand}
\ncmd{\nn}{\nonumber}
\ncmd{\mbf}[1]{\bs{#1}}
\ncmd{\gam}{\gamma}
\ncmd{\sig}{\sigma}
\ncmd{\pha}{\alpha}
\ncmd{\lam}{\lambda}
\ncmd{\dl}{\delta}
\ncmd{\kap}{\kappa}
\ncmd{\Lam}{\Lambda}
\ncmd{\Gam}{\Gamma}
\ncmd{\Dl}{\Delta}
\ncmd{\Ups}{\Upsilon}
\ncmd{\Om}{\Omega}
\ncmd{\veps}{\varepsilon}
\ncmd{\vphi}{\varphi}
\ncmd{\vtheta}{\vartheta}
\ncmd{\tw}{\text{w}}
\ncmd{\pll}{\parallel}
\ncmd{\mc}{\mathcal}
\ncmd{\mf}{\mathfrak}
\ncmd{\bs}{\boldsymbol}
\ncmd{\trans}[1]{{#1}^\intercal}
\ncmd{\note}[1]{{\color{red} {\ding{168} [#1]}}}
\ncmd{\eq}[1]{Eq. \eqref{#1}}
\ncmd{\fig}[1]{Fig. \ref{#1}}
\ncmd{\sm}{\note{`Supplementary Information'}}
\ncmd{\sur}[1]{{\color{blue}{ #1}}}
\ncmd{\new}[1]{{\color{purple}{#1}}}

\begin{document}

\title{Dipolar Weyl semimetals}
\author{Alexander C. Tyner$^{1}$}
\author{Shouvik Sur$^{2}$}
\affiliation{$^{1}$ Graduate Program in Applied Physics, Northwestern University, Evanston, IL 60208}
\affiliation{$^{2}$ Department of Physics and Astronomy, Rice University, Houston, Texas 77005, USA}

\date{\today}
\begin{abstract}
In  Weyl semimetals, Weyl points act as monopoles and antimonopoles of the Berry curvature, with a monopole-antimonopole pair producing a net zero Berry flux. 
When inversion symmetry is preserved, the two-dimensional (2D) planes that separate a monopole-antimonopole pair of Weyl points carry quantized Berry flux. 
In this work, we introduce a class of symmetry-protected Weyl semimetals which host monopole-antimonopole pairs of Weyl points that generate a dipolar Berry flux. 
Thus, both monopolar and dipolar Berry fluxes coexist in the Brillouin zone, which results in two distinct types of topologically non-trivial planes separating the Weyl points, carrying either a quantized monopolar or a quantized dipolar flux.
We construct a topological invariant -- the staggered Chern number -- to measure the latter, and employ it to topologically distinguish between various Weyl points. 
Finally, through a minimal two-band model, we investigate physical signatures of bulk topology, including surface Fermi arcs, zero-energy hinge states, and response to  insertion of a $\pi$-flux vortex. 
\end{abstract}

\maketitle

\twocolumngrid

\section{Introduction}
Weyl semimetals (WSMs)  constitute the most well-known class of topological semimetals which has guided the exploration  of other topological semimetallic phases~\cite{wan2011,Yang2011,burkov2011,Burkov20112,Fang2012,Halasz2012,Zyuzin2012,Vazifeh2013,Goswami2013,HOSUR2013857,Liu2014,Lv2015,xu2015discovery,yang2015weyl,xu2015experimental,Weng2015,Sun2015,soluyanov2015type,jia2016weyl,Yan2017,armitage2018}.
The fact that band-crossing points or Weyl points (WPs) are monopoles of the Berry curvature is one of the most remarkable features of WSMs~\cite{volovik1987,wan2011}.
Arguably, inversion-symmetric WSMs offer the simplest realization of three-dimensional topological semimetals.
In these systems, the two-dimensional planes separating a pair of WPs,  carrying opposite Berry-monopole charges, are classified as Chern insulators.
The Weyl points, respectively, act as source and sink of the quantized Berry flux passing through the Chern planes.
Furthermore, the edge states supported by each Chern insulating layer stack up to give rise to the chiral Fermi-arc states on the surface  of WSMs.
These notions have been generalized by the discovery of higher-order WSMs, where all topologically non-trivial planes are not Chern insulators, and both Fermi arc surface states and hinge-localized zero modes are present at crystal terminations~\cite{ghorashi2020, Wang2020, xie2020,luo2021}.

Recently, Nelson \emph{et al.}~\cite{nelson2022} have shown that the topological critical point separating  Hopf and ordinary insulators realizes a Berry \emph{dipole}, which {  asymptotically act as both a source and a sink} of Berry curvature.
In contrast to an WP, the net Berry flux penetrating a Gaussian surface (GS) enclosing a Berry dipole vanishes.
Instead, the flux is staggered on the GS, with its sign determined by the orientation of the dipole relative to local normals on the GS. 
Therefore, in a hypothetical semimetal,  hosting at least a pair of Berry dipoles, one may expect that the planes separating the dipoles would be threaded by a staggered or dipolar Berry flux.
Are these ``dipolar planes''   topologically non-trivial, and, importantly, does the notion of Berry dipoles lead to hitherto unexplored classes of WSMs?

\begin{figure}[!t]
\centering
\subfloat[\label{fig:texture}]{%
  \includegraphics[width=0.92\columnwidth]{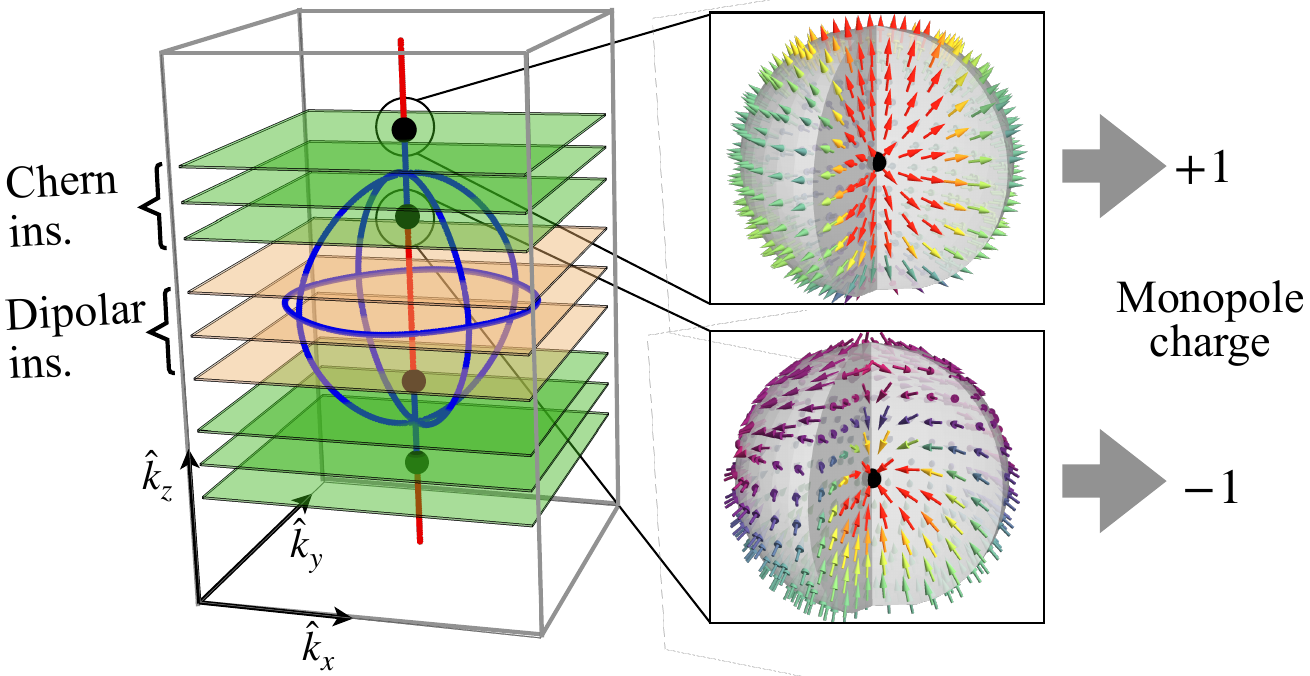}%
}
\hfill
\subfloat[\label{fig:WPoints}]{%
  \includegraphics[width=0.52\columnwidth]{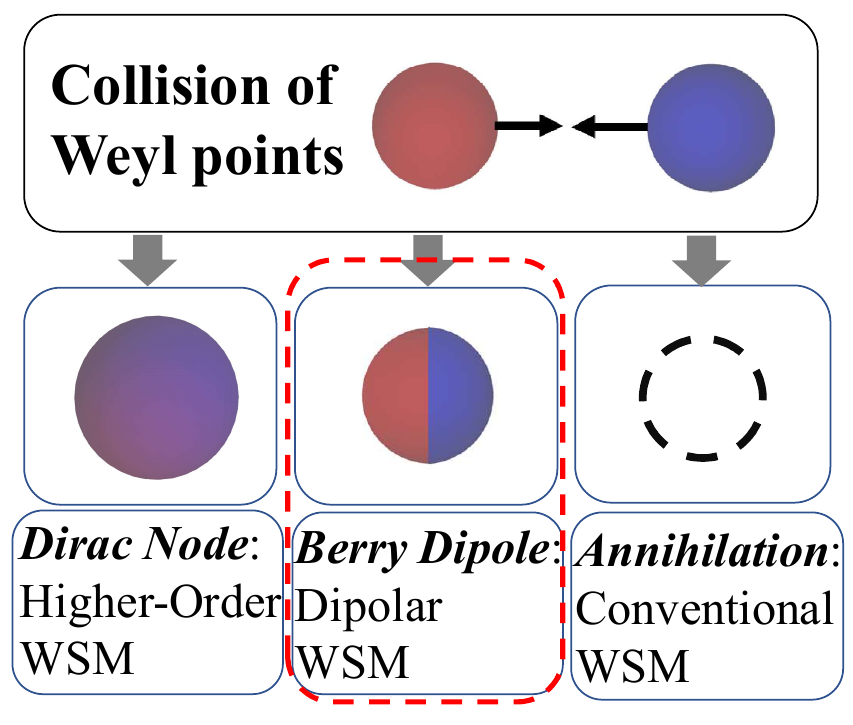}%
}
\hfill
\subfloat[\label{fig:stag-flux}]{%
  \includegraphics[width=0.46\columnwidth]{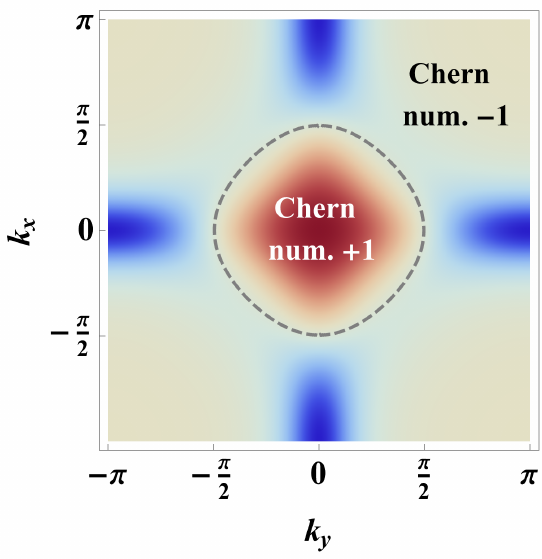}%
}
\caption{Properties of the dipolar Weyl semimetal (WSM)  phase. 
(a) Four Weyl points (WPs) (black spheres) occur on the rotation-axis, with alternating monopole charges. 
The sign is deduced by visualizing the Berry curvature, $\bs {\mc B}$, in the vicinity of the WPs [right].
Depending on their location relative to the WPs, the 2D  planes layered along $\hat k_z$ are either dipolar (red), Chern (green), or ordinary (unmarked) insulators. 
The red [blue] curves indicate points in the Brillouin zone that map to $\hat n = \vec n/|\vec n| = (0, 0, 1)$ [$(0, 0, -1)$], where $\hat n$ is defined in \eq{eq:h}. { The intersection of these curves along the rotation axis marks the locations of band-inversion (black spheres).}
(b) In dipolar- (higher-order~\cite{ghorashi2020})  WSMs, there exist pairs of WPs with monopole-charge $\pm 1$ whose collision leads to a Berry dipole (Dirac point). 
In conventional WSMs, any monopole-antimonopole pair of WPs annihilate upon collision.
(c) On the dipolar insulating planes the distribution of  $\mc B_z$ (color profile)  is such that two Chern insulators, with opposite Chern numbers,  are embedded within the same plane.
In Eq.~\eqref{eq:stagg}, we define the notion of  ``staggered Chern number'' to succinctly capture this pattern.
}
\label{fig:howsm1}
\end{figure}

In this Letter, we introduce a  class of symmetry-protected WSMs, coined ``dipolar Weyl semimetals'',  where Chern, dipolar, and ordinary insulating planes coexist, as summarized in Fig.~\ref{fig:texture}.
The WPs in such WSMs result from  splitting Berry dipoles, which distinguishes dipolar WSMs from both conventional and higher-order WSMs (see Fig.~\ref{fig:WPoints}).
We construct two band models for describing dipolar WSMs, and show that dipolar planes support a quantized, but staggered, Berry flux, as exemplified by Fig.~\ref{fig:stag-flux}.
Therefore, the WPs occuring at the boundary between dipolar and Chern insulators are distinguished from that at the boundary between Chern and 
 ordinary insulators.
Remarkably, on surface-terminations perpendicular to the separation between WPs,  both chiral Fermi arc states and hinge-localized zero-modes are realized.
In order to characterize the bulk topology, we determine the response of the bulk states to  $\pi$-flux vortex insertions, and applied magnetic fields.
In spite of its similarity to  higher-order WSMs~\cite{ghorashi2020,Wang2020}, we note that dipolar WSMs require only  two bands, such that its bulk topology is completely determined by the bands that cross at the WPs. 

\section{Model and phase diagram}
We consider two-band models of WSMs protected by a combination of  $n$-fold improper rotational (or roto-reflection) symmetry $\mc S_m^{z} = \mc{M}_z \circ \mc C_m^{z}$, and two anti-unitary mirror symmetries ($\mc M_1$ and $\mc M_2$). 
Here, $\mc M_z$ is the mirror operator that sends $z \to - z$, and $\mc C_m^z$ generates $m$-fold rotations about the $\hat z$-axis.
We begin with the single-particle Hamiltonian
\begin{align}
H(\bs k) = \sum_{j=1}^3 n_j(\bs k) \sigma_j,
\label{eq:h}
\end{align}
where $\sig_j$ is the $j$-th Pauli matrix, $n_1 = 2 (u_4 u_1 + u_3 u_2)$, $n_2 = 2(u_4 u_2 - u_3 u_1)$, and $n_3 =  u_4^2 + u_3^2 - u_1^2 - u_2^2$.
We require $(u_1, u_2)$  [$u_3$ and $u_4$] to transform under an $E$ [$B$ and $A$, respectively] representation of $\mc C_m^z$, and $\{u_1, u_2, u_4\}$ [$u_3$] to be even [odd]  under $\mc M_z$.
Consequently, $(n_1, n_2)$ [$n_3$] transform under an E [A] representation of $\mc S_m^{z}$.
Since $n_j$'s do not yet have the most general symmetry-allowed form, we modify them as $n_j \to n_j + \nu_j$, where $\nu_j$'s  will be considered as symmetry-allowed perturbations.
Under $\mc S_m^z$, the Hamiltonian transforms covariantly, 
\begin{align}
\mc S_m^z H(\bs k) (\mc S_m^z)^{-1} &= H(\bs k')= U^\dagger H(\bs k) U,
\end{align}
where $\bs k'=(\mf R_m \bs k_\perp, -k_z)$ with $\bs k_\perp = (k_x, k_y)$, $\mf R_m$ implements $m$-fold rotation about $\hat k_z$, and  $U = \exp{-\frac{i\pi}{m} \sig_3}$.
Therefore, $[H(\bs k), \mc S_m^z] = 0$ at high-symmetry points (HSPs) that  satisfy  $\bs k' \equiv \bs k$.
Because of the constraints placed  on $u_j$'s and $\nu_j$'s by $\mc S_m^z$, {$n_{j=1, 2}$ must vanish both at the HSPs and along the rotation-axis}.
The remaining component, $n_3$, is finite in general, and it may change sign between a pair of HSPs only if 
{$u_4^2 + \nu_3$
does not have a fixed sign throughout the Brillouin zone (BZ)}~\footnote{The key difference between our construction and those applied to rotation-symmetry preserving Hopf insulators~\cite{moore2008,nelson2022} is the requirement that $u_3$ transform under the B-representation of $\mc C_m^z$. Requiring the latter forces $u_3 = 0$ on the rotation axis. Consequently, the condition for band-crossing, $u_4^2 = |\nu_3|$, is satisfied without fine-tuning.}.
{While the roto-reflection symmetry guarantees a parameter window where the bands cross along the rotation axis, the anti-unitary mirror symmetries protect  quantized 1D  polarizations   along orthogonal high-symmetry axes.
}

For concreteness,  we focus on $m=4$, and choose  
\begin{align}
& \frac{u_1}{t_p} = \sin{k_x}; ~~
\frac{u_2}{t_p} = \sin{k_y}; ~~
\frac{u_3}{t_d} = \sin{k_z} (\cos{k_y} - \cos{k_x}) \nn \\
& u_4 = t_s\{\Delta - (\cos{k_x} + \cos{k_y} + \gamma \cos{k_z}) \}; \nn \\
& v_1 = 2t_s t_p \delta_\perp \sin{k_x}; ~~
v_2 = 2t_s t_p \delta_\perp \sin{k_y};
v_3 =  -t_s^2 \delta_3^2.
\label{eq:n}
\end{align}
Here, $\{t_p, t_d, t_s, \Delta, \delta_\perp, \delta_z, \gamma \}$ are model parameters.
We note that, at the critical points obtained by setting all $\nu_j = 0$, the Hamiltonian reduces to a form that 
may be obtained from a model of Dirac semimetal (DSM) with a $\mathbb Z_2$-chiral symmetry, $h(\mathbf{k})=\vec u \cdot \vec \Gamma$ with $\Gamma_j$'s being a set of four mutually anti-commuting matrices, by
following the ``Hopf mapping''  procedure in Ref.~\cite{moore2008}.
{The anti-unitary mirror symmetries act as $\mc M_1 H(\bs k) \mc M_1^{-1} \equiv H^*(-k_+, k_-, k_z) = e^{-i\frac{\pi}{4}\sig_3} H(\bs k) e^{i\frac{\pi}{4}\sig_3}$ and $\mc M_2 H(\bs k) \mc M_2^{-1} \equiv H^*(k_+, -k_-, k_z) = e^{i\frac{\pi}{4}\sig_3} H(\bs k) e^{-i\frac{\pi}{4}\sig_3}$, where $k_\pm = \frac{1}{\sqrt{2}}(k_x \pm k_y)$.
}
For simplicity, henceforth, we set $(\gamma, \dl_\perp) = (1, 0)$ and $\Delta, \delta_3 \geq 0$, and note that relaxing these constraints does not qualitatively alter our conclusions.

\begin{figure}[!t]
\centering
\subfloat[\label{fig:phases}]{%
  \includegraphics[width=0.45\columnwidth]{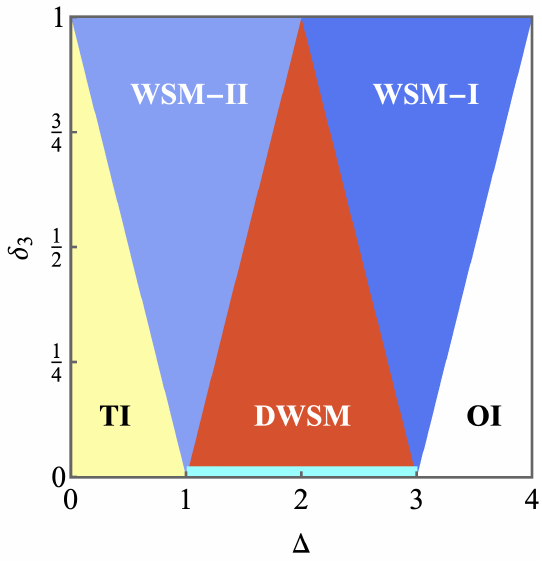}%
}
\hfill
\subfloat[\label{fig:bands}]{%
  \includegraphics[width=0.5\columnwidth]{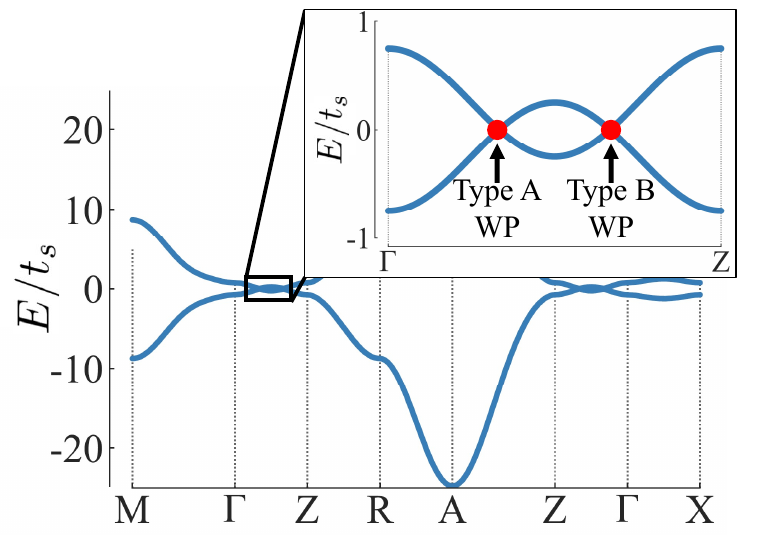}%
}
\hfill
\subfloat[\label{fig:chern}]{%
\includegraphics[width=0.7\columnwidth]{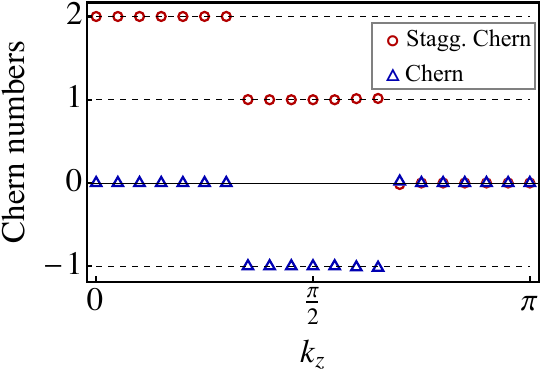}%
}
\caption{Characterization of the phases supported by $H(\bs k)$. 
(a) Phase diagram. 
A dipolar Weyl semimetal (`DWSM') phase exists in the range $(1+\dl_3)< \Delta < (3-\dl_3)$, while conventional WSM  phases (`WSM-I' and `WSM-II') are found when $\delta_3 > |\Dl - 1|$ and $\delta_3 > |\Dl - 3|$.
Here, `TI' (`OI') indicate topological (ordinary) insulating phases.
{  At $\dl_3 = 0$ and $1 < \Dl < 3$ the Weyl points in the DWSM merge on either side of the $k_z$ axis, thereby forming Berry dipoles [c.f. Fig.~\ref{fig:howsm1}(b)]. These critical points are marked in cyan.}
(b) Band structure in the DWSM phase along high-symmetry paths in a tetragonal Brillouin zone. 
Four Weyl points occur along the $\Gamma-Z$ axis. These are classified into two categories (`A' and `B') based on the type of 2D insulating planes they separate [see (c)].
(c) Distribution of the Chern and staggered Chern numbers carried by $(k_x, k_y)$-planes,  as a function of $k_z$.
}
\label{fig:howsm}
\end{figure}

In the BZ of a tetragonal lattice, $[H, \mc S_4^z] = 0$   at $\Gamma \equiv (0,0,0), ~Z \equiv (0, 0, \pi),  ~M \equiv (\pi,\pi, 0)$, and $A \equiv (\pi,\pi, \pi)$ points.
Bands invert  with respect to $\mc S_4^z$ around the $\Gamma$ ($Z$) point for $\delta_3 > |\Delta - 3|$ ($\delta_3 > |\Delta - 1|$).
As shown in Fig.~\ref{fig:phases}, in these regions conventional Weyl semimetallic phases are realized.
Since $n_3$ obtains the same sign at \emph{all} HSPs for $(1+\dl_3) < \Dl < (3-\dl_3)$, no band-inversion would be detected by comparing the eigenvalues of $\mc S_4^z$ at  these points.
Thus, it may appear that in this parameter regime the system is topologically trivial.
This is false, however, because {  along the $\Gamma-Z$ line we identify \emph{two} locations of band-inversion with respect to $\mathcal S_4^z$ at non-HSPs, as shown in Fig~\ref{fig:texture}. 
We refer to these points as ``hidden" band inversions, since their existence cannot be deduced by consulting the
HSPs alone.}
What are the ramifications of such ``hidden'' band inversions?

\section{Dipolar Weyl semimetal}
{In $d$-dimensions, non-trivial topology can be deduced from the texture of $\hat  n(\bs k)$.
In particular, $\hat  n(\bs k)$ maps the  $d$-dimensional BZ  ($T^{d}$) to the 2-sphere ($S^2$) as a function of $\bs k$. 
Valuable insights are obtained by consulting the set of points in the BZ which get mapped to a single point on $S^2$ by $\hat n(\bs k)$ [i.e. the preimages $\hat n(\bs k)$].
In analogy to Hopf insulators~\cite{moore2008}, the preimages of individual points on $S^2$ are 1D curves in the three-dimensional BZ for the dipolar WSMs.  
In contrast to Hopf insulators, however, the preimages of two distinct points on $S^2$ are not necessarily linked in dipolar WSMs.
}
Thus, the linking number or Hopf invariant vanishes.
{In the `DWSM' phase}, the preimages of the ``north'' (red) and ``south'' (blue) poles of $S^2$ -- defined by the simultaneous vanishing of $n_1$ and $n_2$ with $\hat n_3 = \pm 1$, respectively  --   intersect  at $k_z = \pm \cos^{-1}{(\Delta - 2 \pm  \delta_3)}$ along the $\Gamma - Z$ line, as shown in Fig.~\ref{fig:texture}.
{Since $H(\bs k)$ commutes with $\mc S_4^z$ on the polar-preimages, these intersections are locations of band-inversions with respect to $\mc S_4^{z}$}.
The existence of the intersections  is symmetry protected, and remains robust against $\dl_\perp$ as long as $|\dl_\perp| < \delta_3$.
Because  $\vec n(\bs k) = \vec 0$ at the intersection of
the preimages of the north and south poles, these locations correspond to the Weyl points  (see Fig.~\ref{fig:bands}).

\begin{figure}[!t]
\centering
\subfloat[\label{fig:surface1}]{%
  \includegraphics[width=0.49\columnwidth]{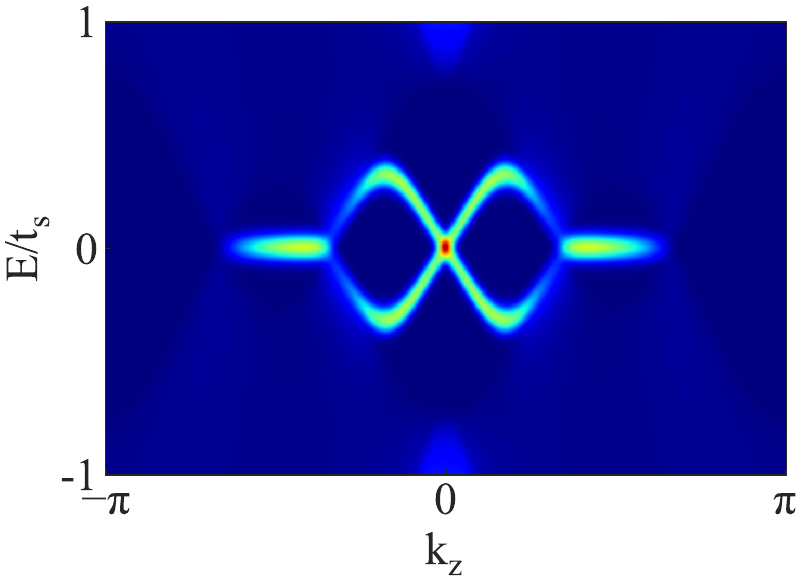}}
\hfill
\subfloat[\label{fig:001surface}]{%
  \includegraphics[width=0.49\columnwidth]{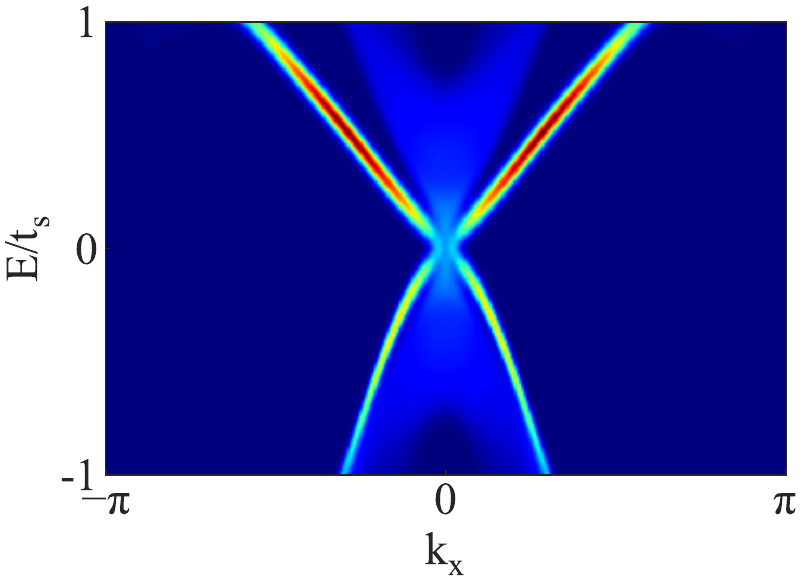}%
}
\vfill
\subfloat[\label{fig:surface2}]{%
  \includegraphics[width=0.49\columnwidth]{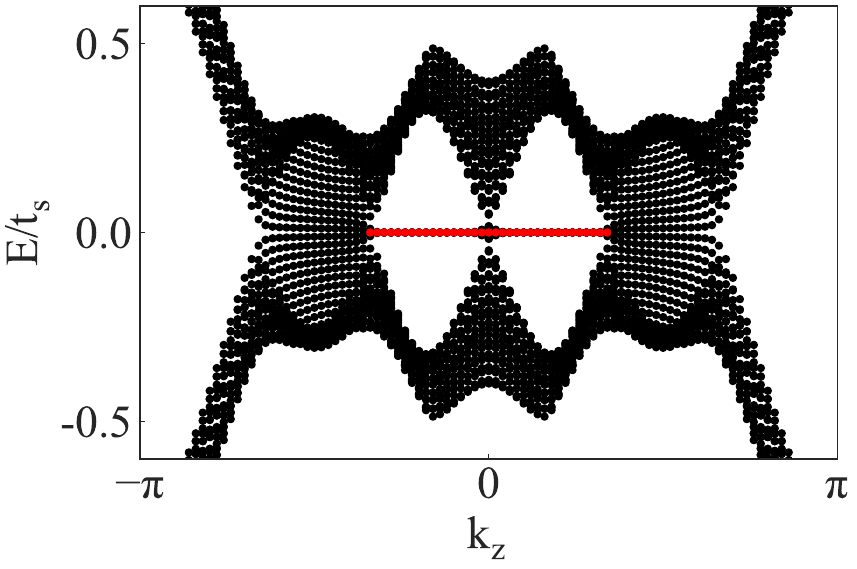}%
}
\hfill
\subfloat[\label{fig:surfaceloc}]{%
  \includegraphics[width=0.49\columnwidth]{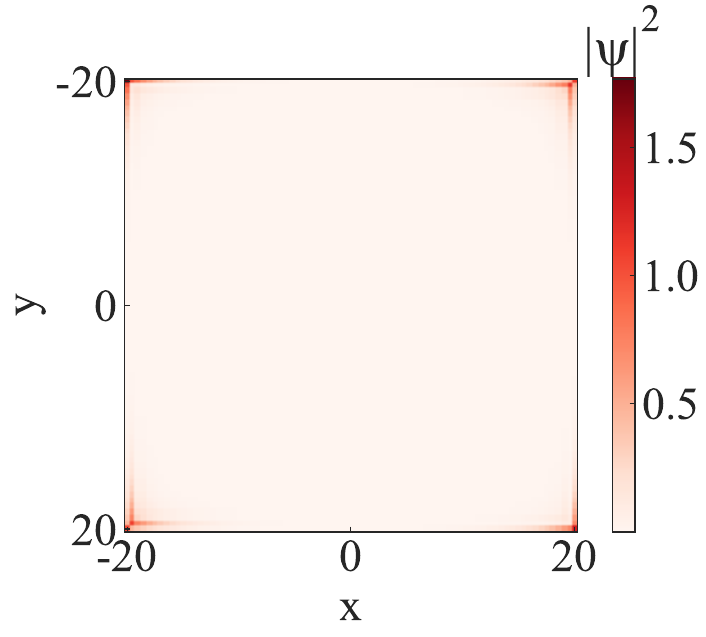}%
}
\caption{Spectra of surface and hinge localized states. 
(a) Spectral density on the (100) surface as a function of $k_{z}$. 
Zero energy Fermi arcs are present between projections of the Weyl-points on the same side of the $k_z$ axis. 
The central region supports a dispersive, non-degenerate, {Dirac cone-like feature} centered at the $\bar \Gam$-point of the surface Brillouin zone. 
(b) Spectra on (001) surface as a function of $k_{x}$ fixing $k_{y}=0$, displaying {  a Dirac cone without a band-crossing point.}  
(c) Spectrum from exact diagonalization with periodic boundary condition only along $\hat z$. 
The dispersionless mid-gap states (marked in red) are localized at the corners of the $(x, y)$-planes, shown in (d), thereby forming dispersionless hinge modes.}
\label{fig:surface}
\end{figure}

\subsection{Topology of bulk states}
By enclosing the band-crossing points by GSs, we determine the net Berry flux  emanating from these band singularities to be $2\pi$ up to an overall sign, as illustrated in Fig.~\ref{fig:texture}.
Therefore, the band-crossing points are unit-strength monopoles of the Berry curvature, and we identify them as WPs. 
When a GS encloses both WPs on a fixed side of the $k_z$-axis, the net Berry flux passing through this surface vanishes.
The region within the GS is topologically non-trivial, however, as a non-zero net Berry \emph{dipole-flux} pierces the surface.
The existence of the dipole-flux can be understood by appealing to the topological critical point at $(\dl_3, \dl_\perp) = (0, 0)$: as $\dl_3 \to 0$, the pair of WPs at $|k_z| = \cos^{-1}{\qty(\Delta -2 \pm \dl_3)}$ collide to yield  a pair of band singularities at { $|k_z| = \cos^{-1}{(\Delta-2)}$}, which act as sources of the dipole-flux (i.e. Berry dipoles; see Fig.~\ref{fig:WPoints}).  
On an infinitesimal GS enclosing the latter  band-crossing points, the Berry flux obtains a staggered form  with the northern (southern) hemisphere supporting a $2\pi$ (-$2\pi$) net flux.
Thus, for $\dl_3>0$, when a Gaussian surface encloses the monopole-antimonopole pair at $|k_z| = \cos^{-1}{\qty(\Delta -2 \pm \dl_3)}$, a net  dipole-flux survives.
This unusual behavior of the WPs in dipolar WSMs is summarized with further details in section II.D. of the `Supplemental Materials'/appendices (SM)~\cite{sm}.
The presence of both monopole and dipole fluxes indicates that topologically non-trivial  $k_z$ planes in the BZ are not limited to being 2D Chern insulators.

{  Indeed, the $k_z$-planes between $k_z = \pm \cos^{-1}{(\Dl - 2 + \dl_3)}$ (red planes in Fig.~\ref{fig:texture}) support a dipolar version of the Berry flux which reflects the presence of a skyrmionium texture~\cite{skyrmonium} for $\vec n(\bs k)$.
Skyrmioniums are composed of two  oppositely charged but non-overlapping skyrmions; consequently, they do not support a finite Chern number. 
It is possible to define a quantized topological invariant, however, that distinguishes a skyrmionium-carrying plane from an ordinary 2D insulator.
}
In order to emphasize its origin in a staggered distribution of Berry curvature on the $\bs k_\perp$ plane, we call this  topological invariant  ``staggered Chern number'', and define it as 
\begin{align}
\mf C_{\text{stagg}}(k_z) = \frac{1}{2\pi} \int \dd{\bs k_\perp} \mc B_z(\bs k_\perp, k_z) f_{\text{stagg}}(\bs k_\perp),
\label{eq:stagg}
\end{align}
where $\mc B_z(\bs k_\perp, k_z)$ is the $\hat k_z$ component of the Berry curvature, and  $f_{\text{stagg}}(\bs k_\perp)$ is a weight function that is determined by $H(\bs k)$ such that $\int \dd{\bs k_\perp} f_{\text{stagg}}(\bs k_\perp) = 0$.
Since $\mf C_{\text{stagg}}$ is effectively the difference of Chern numbers of the  same magnitude but opposite signs,  it equals the Chern number when the staggering is absent.
If a quantized staggering of flux is present, then $\mf C_{\text{stagg}}$ will be an even integer, in analogy to familiar mirror~\cite{Teo2008} or spin~\cite{bernevig2006quantum} Chern numbers.
While on mirror- or spin-Chern number carrying planes the Chern number is staggered in an internal sub-space, here, it is staggered in the momentum space.
An alternative formulation of $\mf C_{\text{stagg}}$ that does not require an explicit knowledge of $f_{\text{stagg}}$, but utilizes effective Su-Schrieffer-Heeger~\cite{ssh} forms of the Hamiltonian on the mirror axes, is provided in section II.B of the SM~\cite{sm}.

We plot $\mf C_{\text{stagg}}$ as a function of $k_z$ in Fig.~\ref{fig:chern}, and    relegate the details of the calculation to the SM~\cite{sm}, along with the explicit form of $f_{\text{stagg}}(\bs k_\perp)$.
We obtain $\mf C_{\text{stagg}} = 2$ for all planes with $|k_z| < \cos^{-1}{(\Dl - 2 + \dl_3)}$, indicating intertwining of two regions with opposite Chern numbers.
We note that the $k_z = 0$ and $\pi$ planes of the dipolar WSM  are identical to respective planes in the Moore-Ran-Wen class of models of Hopf insulators~\cite{moore2008}.
Consequently, \emph{these special planes in Hopf insulators are also characterized by a quantized  staggered-Chern number}.
The staggering of the Chern number in dipolar WSMs, as well as Hopf insulators, originates from non-trivial 2D winding numbers supported by corresponding $\bs k_\perp$ planes of the system {  governed by  $h(\bs k) = \vec u \cdot \vec \Gamma$.} 
{  In the present case, $h(\bs k)$ describes a $\mathbb{Z}_2$-chiral DSM with a pair of band-crossings along the $k_z$-axis.
The $k_z$-planes lying between the Dirac points are characterized by the relative-Chern number~\cite{tyner2020}, which is an $\mathbb Z$-valued 2D bulk invariant and a variant of the spin-Chern number, {and it is protected by both bulk- and spin-gaps}~\cite{Prodan2009,Lin2022Spin}. 
The non-trivial $\mf C_{\text{stagg}}$ is thus a manifestation of the non-trivial relative- or spin-Chern number supported by the $k_z$-planes of chiral DSMs~\cite{tyner2020}.}
Thus, under the Hopf map, \emph{the staggering of the Chern number in a Kramers degenerate subspace maps to its staggering in momentum space}.

The $k_z$-planes separating WPs on the \emph{same} side of the $k_z$-axis, i.e. $\cos^{-1}{\qty(\Delta -2 + \dl_3)} < |k_z| < \cos^{-1}{\qty(\Delta -2 - \dl_3)}$ (green planes in Fig.~\ref{fig:texture}), carry  a Chern number $\mf C = -1$, as shown in Fig.~\ref{fig:chern}. 
By contrast, all planes with $|k_z| > \cos^{-1}{\qty(\Delta -2 - \dl_3)}$  are topologically trivial with $\mf C_{\text{stagg}} = 0 = \mf C$.
Thus, the $k_z$-plane hosting a  type-A [-B] WP (see inset in  Fig.~\ref{fig:bands}) may be interpreted as the topological critical point separating a staggered-Chern and a Chern insulator [a Chern and an ordinary insulator]. 
{  Here, while the $k_z$-plane hosting a type-B WP supports a half-integer Chern and staggered-Chern number of equal magnitude, the $k_z$-plane hosting type-A WP supports a half integer Chern number and a \emph{distinct}  half-integer staggered Chern number~\cite{sm}.}

\subsection{Surface and hinge states}
{The states on the (100) and (010) surfaces are sensitive to the texture of $\hat n(\bs k)$ in the bulk.
} 
As a representative example,  we portray the topologically protected states on the (100) surface in Fig.~\ref{fig:surface1}.
Fermi arcs are found to connect the projections of WPs on the same side of the $k_z$ axis, indicating their origin in the non-trivial Chern insulating planes.
The staggered-Chern or dipolar  planes have a distinct topological response to surface terminations, which is reminiscent of  $sp$-Dirac semimetals~\cite{tyner2020, wei2021}.
In particular, generic dipolar  planes support a pair of gapped edge states.
On the  $k_z = 0$ (and $\pi$)  plane $\mathcal S_4^z$ reduces to $\mathcal C_4^z$, which allows it to support gapless edge states.
This leads to the single copy of Dirac cone centered at the $\bar \Gam$ point  of the surface-BZ. 
The pair of degenerate surface states at the $\bar \Gam$ point is protected by a quantized 1D winding number in the bulk along the $k_x$-axis~\cite{sm}.
The states on the (001) surface exist as long as the $\mc S_4^z$ symmetry is preserved.
As shown in Fig.~\ref{fig:001surface}, these states form a {Dirac cone-like feature} about the zone-center of the (001) surface BZ, with the band-crossing point absent. 
The center of the (001) surface BZ corresponds to the projection of the rotation axis, which accounts for the lack of normalizibilty at this point~\cite{tyner2020}.

While generic $k_z$-planes in the region $|k_z| < \cos^{-1}{(\Dl - 2 + \dl_3)}$ support gapped edge states, they also support corner-localized zero modes, {which are protected by the anti-unitary mirror symmetries.}
These corner localized modes stack along the $\hat z$-direction to give rise to hinge localized zero-modes.
{ 
In Fig.~\ref{fig:surface2} and \ref{fig:surfaceloc} we identify these hinge-states by exactly diagonalizing the Hamiltonian with periodic boundary condition only along $\hat z$. 
In Section III of the SM~\cite{sm} (also see Ref.~\cite{yan2019})  we detail the presence of a quantized, one-dimensional winding number along the diagonal axes, which  protect the corner localized zero-modes in accordance with Ref. \cite{sur2022}. }
Thus,  surface- and hinge-localized Fermi arcs are simultaneously present in  dipolar WSMs, which is reminiscent of higher-order topological semimetals~\cite{ghorashi2020, Wang2020, wieder2020, tyner2020}.

\begin{figure}[!t]
\centering
\subfloat[\label{fig:HWVortex}]{%
  \includegraphics[width=\columnwidth]{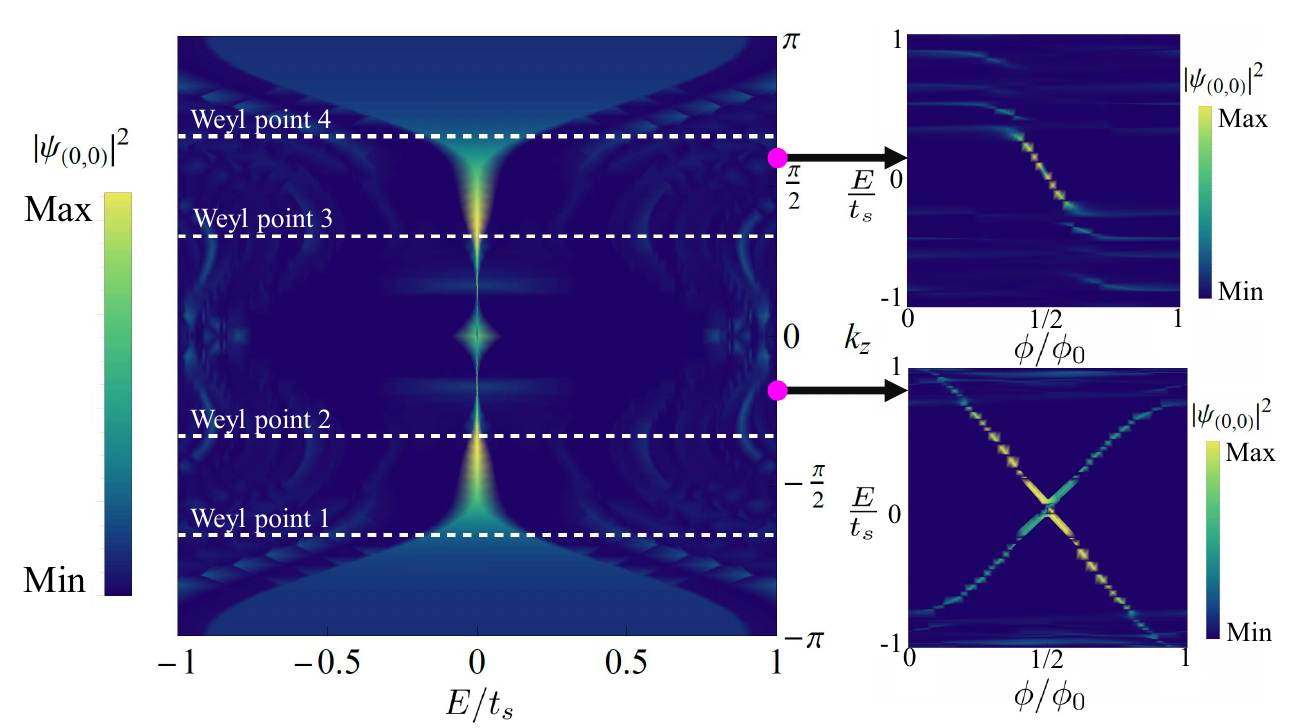}%
}
\vfill
\subfloat[\label{fig:VortexZMs}]{%
  \includegraphics[width=.75\columnwidth]{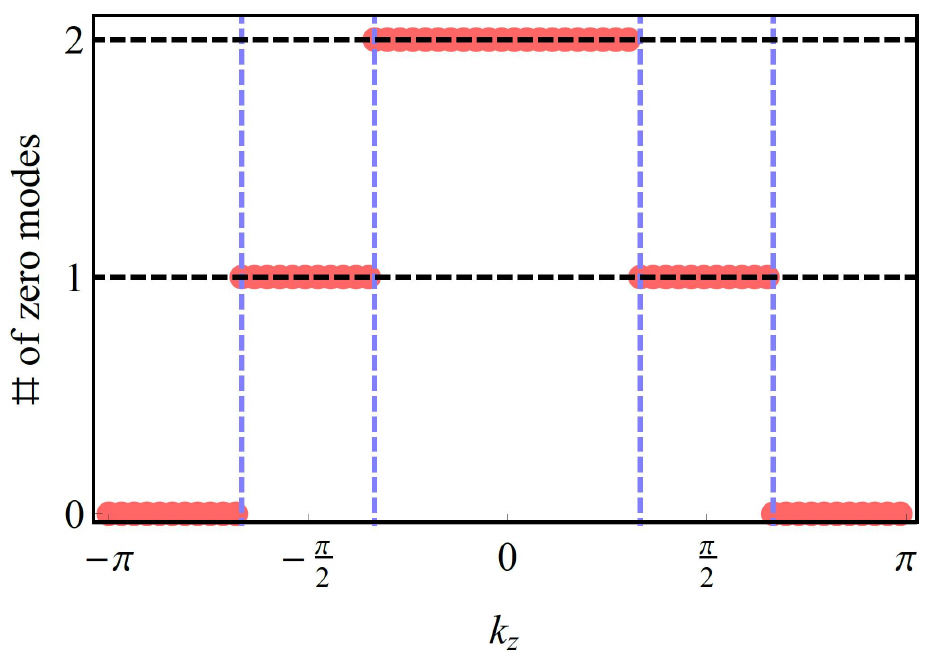}%
}
\caption{Vortex-bound states as a diagnostic of bulk topology. (a) Probability density (PD) at the location of vortex for 120 states closest to zero energy. 
Left: PD for vortex-flux  $\phi=\phi_{0}/2$ as a function of energy and $k_{z}$. 
Here, $\phi_{0}$ is the quantum of flux.
Zero modes exist for \emph{all} topologically non-trivial planes. 
Right: PD as a function of $\phi$ and energy, fixing $k_{z}$. 
Both Chern (top right) and staggered-Chern planes (bottom right) support charge pumping.
(b) Number of vortex-bound zero modes upon insertion of magnetic flux tube with  $\phi=\phi_{0}/2$. 
The distinct number of zero-modes establishes the different topological character of $\bs k_\perp$ planes as a function of $k_{z}$.
} 
\label{fig:Vortex}
\end{figure}

\subsection{Vortex-bound states}
In the position-space, field-theoretical calculations have proven that the bulk topological invariant in the ground state can be computed through insertion of an electromagnetic $\pi$-flux vortex. 
In particular, the number of states bound to the vortex corresponds to the magnitude of the quantized flux in the ground state of a 2D topological insulator\cite{SpinChargeSCZ,VishwanathPiFlux,MESAROS2013977,slager2012,tyner2020,tyner2021quantized}.
Moreover, as the magnitude of flux carried by the vortex is tuned between zero and the flux quanta, $\phi_{0}=h/e$, the  vortex-bound modes are pumped across the bulk band-gap. 

To unambiguously examine the topological response of the two-dimensional planes stacked along $\hat k_{z}$, we consider them as independent layers of 2D insulators, labeled by $k_z$.
In the position-space for each layer, we insert a vortex tube at the origin,  carrying a flux $\phi$.
We exactly diagonalize the resultant Hamiltonian on a $20 \times 20$ lattice, which yields a $k_z$-dependent energy spectrum, as shown in Fig.~\ref{fig:HWVortex}. 
The unit strength Chern planes at $\cos^{-1}{\qty(\Delta -2 + \dl_3)} < |k_z| < \cos^{-1}{\qty(\Delta -2 - \dl_3)}$ support a single vortex-bound state, which is pumped across the bulk band-gap as a function of  $\phi$.
The dipolar  or staggered-Chern planes at $|k_z| < \cos^{-1}{\qty(\Delta -2 + \dl_3)}$ support \emph{two} states at the vortex, each corresponding to a Chern sector.
Since the Chern sectors carry opposite Chern numbers, these vortex-bound states are pumped in opposite directions as a function of $\phi$, reminiscent of spin-Hall insulators\cite{SpinChargeSCZ,VishwanathPiFlux,Wang_2010,tynerbismuthene}.
Importantly, as demonstrated in Fig.~\ref{fig:VortexZMs}, when the strength of flux is held fixed at $\phi=\phi_{0}/2$, the number of vortex-bound modes can be used as a quantized diagnostic of the topology of the 2D layers.
Since both $\mf C_{\text{stagg}}$ and the number of vortex-bound zero modes effectively count the number of Chern sectors in each layer, they have an identical response (c.f.  Figs.~\ref{fig:chern} and ~\ref{fig:VortexZMs}).

\section{Conclusion}
We introduced a class of WSMs where both dipole- and monopole-flux carrying planes are present.
A topological invariant, the staggered Chern number, is formulated for diagnosing the presence of quantized dipolar-flux.
Through flux insertions, the dipolar planes are shown to have a topological response that is analogous to generalized spin-Hall insulators.
With the help of a two-band model, we demonstrated that surface and hinge states are reminiscent of higher order topological semimetals. 
A detailed comparison among dipolar, higher-order, and conventional  WSMs is presented in the SM~\cite{sm} (see Fig.~\ref{fig:comparison}), where we show that the clearest distinction between a dipolar and a higher-order WSM arises in the Landau level spectra.

The two-band model discussed here  follows a similar principle of construction as Hopf-insulators.
Therefore, we expect it would be  possible to simulate it within the same platforms proposed for realizing the latter~\cite{yuan2017,deng2018,yi2019,schuster2019b,schuster2021,unal2019, wang2023}.
Further, it is possible to construct variants of the same model, with potentially more exotic topological singularities~\cite{deng2013Hopf,ezawa2017, graf2022, zhu2023}.
Both considerations are left to future works. 

\acknowledgments{A.C.T. was supported by the start-up funds of Pallab Goswami at Northwestern University and the National Science Foundation MRSEC program (DMR-1720139) at the Materials Research Center of Northwestern University. S.S. was supported in parts by the U.S. Department of Energy Computational Materials Sciences (CMS) program under Award Number DE-SC0020177, and National Science Foundation under Grant No. DMR2220603.
We thank the anonymous referee for helpful comments which has improved the quality of this work.}

\clearpage
\appendix
\onecolumngrid

\renewcommand{\theequation}{S\arabic{equation}}
\renewcommand{\thefigure}{S\arabic{figure}}
\renewcommand{\thesection}{\Roman{section}}
\renewcommand{\thesubsection}{\Alph{subsection}}
\setcounter{figure}{0}

\section{Comparison with other  models of Weyl semimetals with 4 Weyl points} 
In this section we compare the dipolar Weyl semimetal (WSM) with a conventional WSM with four Weyl points and a model of time-reversal symmetry (TRS) broken higher-order WSM.

\subsection{Conventional WSM with 4 Weyl points} \label{app:conventional}
Here, we show that more conventional 2-band models of TRS-broken Weyl semimetals can support $2n$ band crossings points with $n > 1$.
Such models are obtained by introducing further neighbor hoppings into the term that controls  band-inversion,
\begin{align}\label{eq:4wp}
H = t_p \sin{k_x} \sig_1 + t_p \sin{k_y} \sig_2 + t_s(\Delta - \cos{k_x} - \cos{k_y} - \sum_{n=1}^N \alpha_n \cos{n k_z}) \sig_3.
\end{align}
For convenience we set $\pha_1 = 1$, and consider a physical range of parameters, $|\pha_{n>1}| < 1$.

The condition for band-crossings along the $k_z$-axis passing through $(k_x, k_y)= (0,0)$ is 
\begin{align}
\sum_{n=1}^N \alpha_n \cos{n k_z} = \Delta - 2.
\end{align}
The simplest case corresponds to $N=1$, where band crossings occur at 
\begin{align}
k_z = k_{z,\pm} \coloneqq \pm \cos^{-1}(\Dl -2)
\end{align}
Thus, a pair of Weyl points are present as long $1 < \Dl < 3$.

For $N=2$, an additional pair of Weyl points appear in the region
\begin{align}
\qty(2 - \frac{1}{\sqrt{2}}) < \Dl < \qty(2 + \frac{1}{\sqrt{2}}) 
\quad \mbox{and} \quad 
\mbox{sign}(\alpha_c) \alpha_2 > |\alpha_c| \quad
\mbox{with} \quad
\pha_c = \Theta(\Dl - 2) (\Dl - 3) + \Theta(2 - \Dl) (\Dl - 1).
\end{align}
Note that four Weyl points also exist in the region $\frac{1}{\sqrt{2}} < |\Dl - 2| < \frac{3}{4}$, but the constraint on $\alpha_2$ is more complicated.

These WSM phases, however,  support 1st-order Weyl points only.
This is established by the 1D winding number along $k_x$ and $k_x = k_y$ axes vanishing on the $k_z$ planes that do not support  finite Chern numbers.
Therefore, the above model can support only Fermi-arc surface states. 
Thus, both the bulk and boundary signatures are in sharp contrast to dipolar-WSMs.

\subsection{Comparison of WSMs with four Weyl points}\label{modelcomparison} 
Here we provide a more explicit comparison among dipolar WSM, the model in section~\ref{app:conventional}, and TRS-broken higher-order WSM (HOWSM). 
For the HOWSM phase, we select a modified version of HOWSM proposed by Ghorashi  et. al\cite{ghorashi2020}. The model is written as, 
\begin{equation}\label{eq:HOWSM}
    H=t_{p} \sin k_{x} \Gamma_{1} + t_{p} \sin k_{y} \Gamma_{2}+ t_{d}\sin k_{z}(\cos k_{x}-\cos k_{y})\Gamma_{3}+ t_{s}(\Delta+\cos k_{x}+\cos k_{y} + \cos k_{z})\Gamma_{5}+0.4i\Gamma_{1} \Gamma_{2},
\end{equation}
where $t_{s,p,d}$ are constants with units of energy, $\Delta$ is a dimensionless parameter, and the lattice constant has been set to unity. The $\Gamma$ matrices are defined as, $\Gamma_{i=1,2,3}=\tau_{1}\otimes \sigma_{i},\; \Gamma_{4}=\tau_{2}\otimes \sigma_{0},\; \Gamma_{5}=\tau_{3}\sigma_{0}$. The model used by Ghorashi et. al\cite{ghorashi2020}, is recovered by fixing $t_{d}\rightarrow t_{d}/\sin k_{z}$. For the current calculations we fix $t_{p}=t_{s}=t_{d}$ and $\Delta=-2$. 
In addition, we employ a minimal model of a conventional WSM given in eq. \eqref{eq:4wp}, fixing $N=2$, $\alpha_{1}=-\alpha_{2}=1$, and $\Delta=2.5$.

\begin{figure}
    \centering
    \includegraphics[width=12cm]{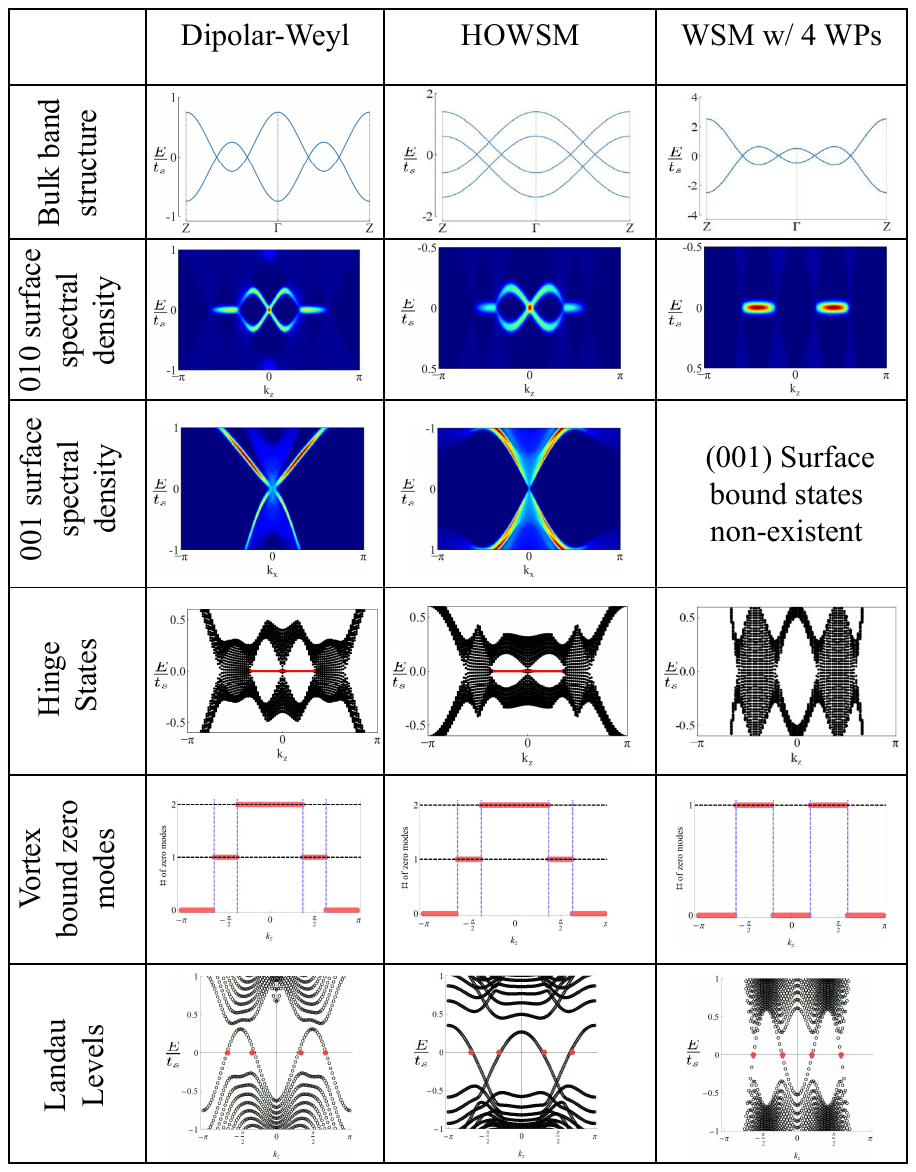}
    \caption{The main energetic and topological features of three distinct models of WSMs with 4 WPs at the Fermi energy are compared: (1) the Dipolar-Weyl model using the same parameters used in the main-body, (2) a modified version of the HOWSM model proposed by Ghorashi et. al\cite{ghorashi2020}, and (3) the model detailed in eq. \eqref{eq:4wp}. While the HOWSM model supports four bands, there is no clear way to distinguish the Dipolar-Weyl model from that of eq. \eqref{eq:4wp} via the bulk band structure. By contrast the HOWSM and Dipolar-Weyl model are indistinguishable when viewing common diagnostics of topological order.}
    \label{fig:comparison}
\end{figure}

Key features of these models are then tabulated in Fig. \eqref{fig:comparison}. The first row of this table displays the bulk band structure for each model. The HOWSM is immediately distinguishable from the dipolar and conventional phases due to the requirement of at least two occupied bands to realize this phase. By contrast, when observing the spectral density on the (010) surface, the HOWSM and dipolar WSM phases appear nearly indistinguishable. Both support both gapped and gapless states, with the $k_{z}$ planes supporting gapped surface states simultaneously admitting four, zero-energy corner-localized modes. The (010) spectral density of the conventional WSM supports only the zero-energy Fermi arcs that are the classic signature of an WSM. 
The (001) surface spectral density  provides a clearer route to differentiate between each model. Conventional WSMs do not support (001) surface bound modes. The HOWSM and dipolar WSM phases support Dirac cones, but the Dirac point itself is absent as it is non-normalizable in the semi-infinite limit. 
Hinge-localized zero modes are present in both dipolar and higher-order WSMs, while such states are absent in the conventional WSM.

As stated in the main body, insertion of an electromagnetic vortex serves as a non-perturbative probe of bulk topological order. We insert a $\pi$-flux vortex into each of the $xy$ planes, maintaining periodic boundary conditions. Examining the number of zero-energy bound states in each of the planes perpendicular to the axis of nodal separation, the HOWSM and dipolar WSM display an identical response. The planes in which the Vortex binds a single zero mode can be understood through correspondence with the non-zero bulk Chern number. On the other hand, the planes supporting two zero-energy vortex bound modes display identical response stemming from distinct bulk invariants, the relative Chern number and staggered Chern number for the HOWSM and dipolar WSM respectively.

Therefore, we find that common diagnostics of topological order in weakly interacting systems cannot distinguish between the TRS-broken HOWSM and dipolar-WSM.

The TRS-broken HOWSM and dipolar-WSM states can be distinguished, however, by comparing the respective Landau level spectra.
We compute the LL spectrum in a magnetic field applied along the $\hat{z}$-axis. We consider the full lattice model and employ units where the electric charge and the lattice constant are set to unity. In these units and with the gauge choice, $\mathbf{A}=(0,Bx,0)$,  $k_{y}$ is dimensionless and $B=2m\pi/L_{x}$ with $m\in \mathbb{Z}$. LLs in the weak field limit ($m=2$) is shown in Fig. \eqref{fig:comparison}. 
Since the minimal model for the dipolar WSM requires only two bands while the HOWSM requires four, the zero modes generated upon application of a magnetic field parallel to the direction of nodal-separation give rise to one and two bands, respectively. 

\section{Monopole and Dipole flux} \label{app:flux} 
Here we present the details for the computation of monopole and dipole flux, both in the vicinity of the Weyl points, as well as the $k_z$ planes.

\subsection{Generic $k_z$ planes}
In order to compute the staggered Chern number for generic values of $\delta_3$ and $k_z$, we must be able to identify the boundary which separates the skyrmion and anti-skyrmion texture which are superimposed to produce the skyrmionium texture\cite{skyrmonium} which is characteristic of an insulator supporting a staggered Chern number. 
The presence of a skyrmionium texture is protected by the topology of the chiral Dirac semimetal, from which a Hopf map is implemented to construct the dipolar Weyl Hamiltonian. However, the form of the skyrmionium texture is model dependent, thus the  boundary between skyrmion and anti-skyrmion is a non-local quantity in the momentum space that is complicated to identify in general. Nonetheless, the pattern of flux-assisted charge pumping can be used as a universal probe to diagnose a dipolar insulating plane as discussed in section III.C of the main text.

For computing the staggered-Chern number at generic $\delta_3$ and $k_z$ we change basis, $k_\pm = (k_x \pm k_y)/\sqrt{2}$, and introduce $(n_-, n_+) = \frac{1}{\sqrt{2}}(n_2 - n_1, n_2 + n_1)$.
Thus, up to a global unitary transformation, the Hamiltonian obtains the form 
\begin{align}
H(k_+, k_-, k_z) = n_+(k_+, k_-, k_z) \sigma_1 + n_-(k_+, k_-, k_z) \sigma_2 + n_3(k_+, k_-, k_z) \sigma_3.
\end{align}
It can be easily checked that $H$ is invariant under $k_\pm \to k_\pm + 2\sqrt{2} \pi$.
By defining $\vec N = (n_+, n_-, n_3)$ and $\hat N = \vec N/|\vec N|$, we compute the Berry curvature density on the $k_z$ planes,
\begin{align}
\mc B_z(k_+, k_-, k_z) = \frac{1}{4} \hat{N}\cdot(\partial_{k_+} \hat N \times \partial_{k_-} \hat N).
\end{align}
Since the area of the Brillouin zone (BZ) for $k_\pm$ is twice that of $(k_x, k_y)$, we have  introduced an extra multiplicative factor to halve the net flux through the enlarged BZ, which yields the physical result.
The weight function is determined with the help of $n_+$,
\begin{align}
f_{\text{stagg}}(k_+, k_-, k_z) = & \mbox{sign}\Biggl[\qty{
\cos \left(\frac{k_-}{\sqrt{2}}\right) \qty(1 -t_d \sin{k_z} + \cos \left(\sqrt{2} k_+\right) \left(t_d \sin {k_z}+1\right)) - \cos \left(\frac{k_+}{\sqrt{2}}\right) \left(\delta +\Delta -\cos{k_z} \right)
} \nn \\
& \qquad\qquad \times \qty{\Theta\qty(\cos{\frac{k_+}{\sqrt{2}}}) \Theta\qty(\cos{\frac{k_-}{\sqrt{2}}}) - \frac{1}{2}
} \Biggr]
\end{align}

\begin{figure}[!t]
\centering 
\subfloat[]{
\includegraphics[width=0.35\columnwidth]{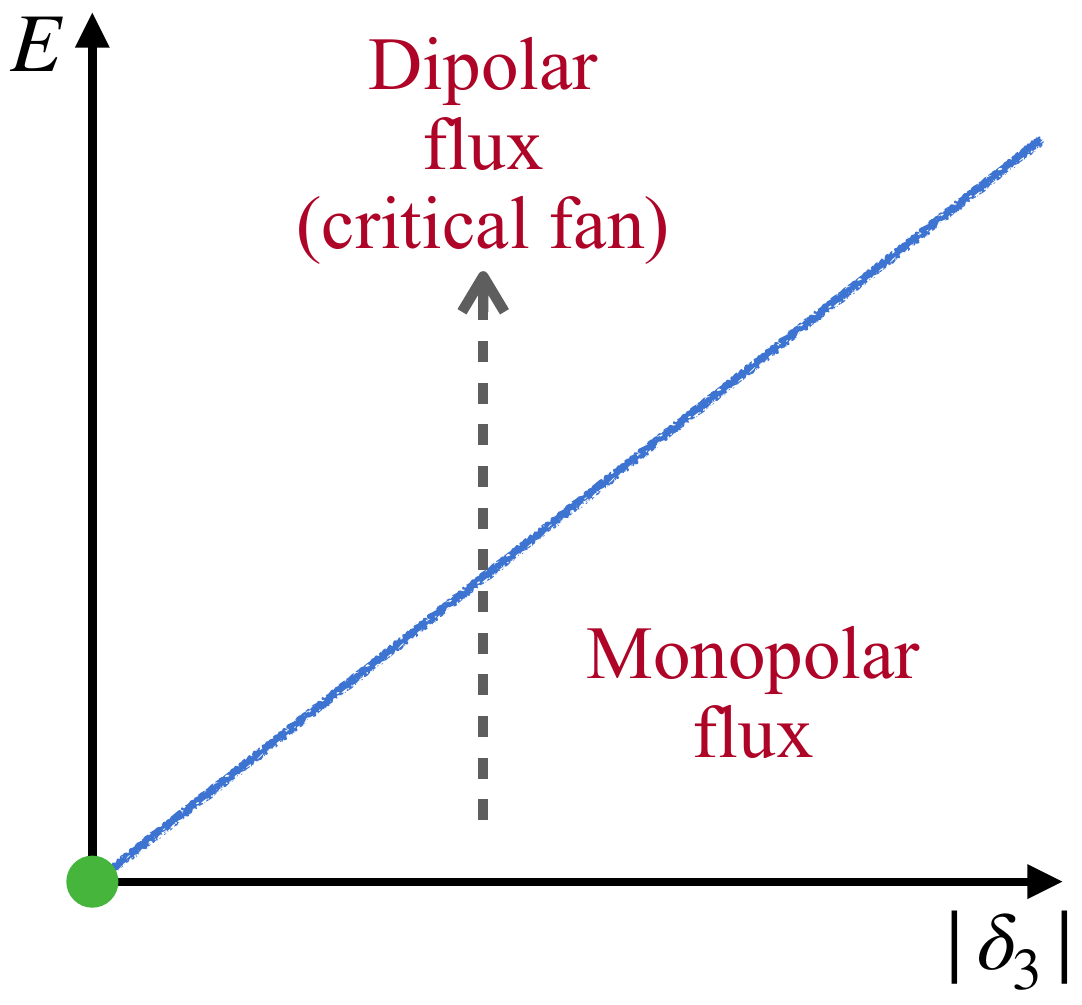}%
}
\hfill
\subfloat[]{
\includegraphics[width=0.5\columnwidth]{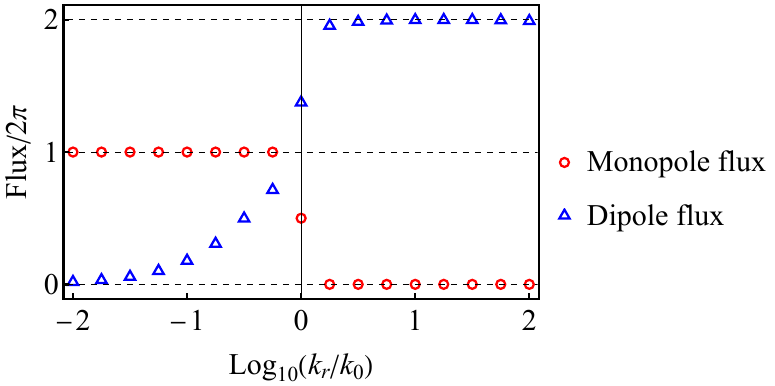}%
}
\caption{Change in the nature of flux as a function of radius of the Gaussian sphere centered at the Weyl point at $k_z = \cos^{-1}(\Delta - 2 - \delta_3)$. 
(a) Expected crossover in the behavior of Berry flux as measured on the Gaussian surface in the dipolar Weyl semimetal phase. 
The vertical axis represents an  energy/momentum scale, and the green dot locates the critical point where the Weyl points merge to form a Berry dipole.
(b) Actual behavior of the Berry flux obtained by numerically computing the Berry fluxes, as detailed in section II.C. 
Here, $k_0 = \qty[\cos^{-1}(\Delta - 2 - \delta_3) - \cos^{-1}(\Delta - 2 + \delta_3)]$ is the distance between the two Weyl points on the same side of $k_z$ axis.
}
\label{fig:crossover}
\end{figure}

\subsection{Formulation of $\mf C_{\text{stagg}}$ independent of $f_{\text{stagg}}$}   
The existence of {both  skyrmion and skyrmionium textures} in the presence of two {distinct} anti-unitary mirror symmetries ($\mc M_1$ and $\mc M_2$) gives rise to quantized one-dimensional winding numbers along orthogonal high-symmetry lines. For the model in eq. 6 of the main body, these are the $k_x =\pm k_y$ lines. Along these high-symmetry axes the Hamiltonian appears as a non-trivial Su-Schrieffer-Heeger (SSH) model\cite{ssh} with winding number 1 (2) in planes supporting monopole(dipole) Berry flux. For generality, we will refer to these high-symmetry lines as $k_{1,2}$. 
\par 
Let us define the Berry gauge connection along $k_{1,2}$, as 
\begin{equation}
    A_{i}(k_{i},k_{z})=-iU(k_{i},k_{z})\partial_{k_{i}}U^{\dagger}(k_{i},k_{z}),
\end{equation}
where $U(\mathbf{k})$ is the diagonalizing matrix for the Hamiltonian, $H(\mathbf{k})=\sum_{j=1}^{3}n_{j}\sigma_{j}$, and it takes the form, 
\begin{equation}
    U(\mathbf{k})= \frac{1}{|\vec n|}[n_{3}\sigma_{0}+i(n_{1}\sigma_{1}+n_{2}\sigma_{2})].
\end{equation}
\par 
If $[A_{1},A_{2}] \neq 0$, then $\vec n = (n_1, n_2, n_3)$ has a non-trivial texture, which can be succinctly captured by 
\begin{equation}\label{eq:staggInd}
      \mf C_{\text{stagg}}=\frac{1}{2\pi}\frac{\iint d^{2}k \text{Tr}\left(S_{12}\cdot[A_{1},A_{2}]\right)\big)}{\left(\big |\iint d^{2}k \text{Tr}\left(S_{12}\cdot[A_{1},A_{2}]\right)\big|\right)^{1/2}},
\end{equation}
 where $S_{12}=S_{1}\cdot S_{2},$ such that $S_{j}$ is the generator of chiral symmetry for the embedded SSH Hamiltonian corresponding to the Berry gauge connection $A_{j}$.
We note that $\mf C_{\text{stagg}}$ reduces to the Chern number when the staggering is absent.

\begin{figure}
    \centering
    \includegraphics[width=8cm]{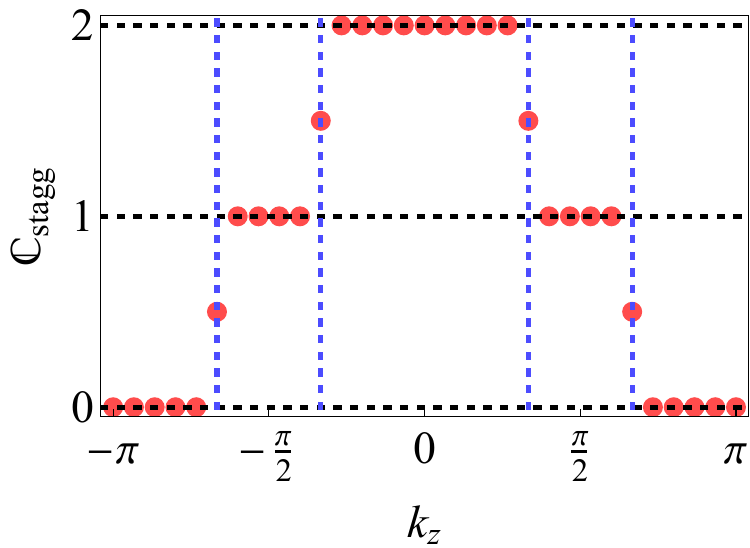}
    \caption{Computation of $\mf C_{\text{stagg}}$ utilizing eq. \eqref{eq:staggInd}. This formulation does not require knowledge of $f_{\text{stagg}}$, but returns identical results.}
    \label{fig:cstaggnew}
\end{figure}

The results of applying Eq. \eqref{eq:staggInd} to the model of a dipolar WSM presented in the main body are shown in Fig. \eqref{fig:cstaggnew}, detailing identical results to those found upon identification of $f_{\text{stagg}}$.

\subsection{Weyl planes}
In a typical Weyl semimetal the plane on which a Weyl point lie (henceforth, called a ``Weyl-plane'') supports a half-integer Chern number, which implies the presence of a meron rather than a skyrmion texture. 

In dipolar Weyl semimetals, the Weyl-plane separating Chern and ordinary insulating planes support a similar meron texture, which result in an half-integer Chern number.
We note that, due of the manner in which the staggered Chern number is defined, these planes will trivially support an identical-in-magnitude staggered Chern invariant.

\begin{table}[h!]
\centering
\begin{tabular}{ |c|c| } 
 \hline
 {\bf Texture} & {\bf Invariant} \\ 
 \hline \hline
 skyrmion & CN $\in \mathbb Z$; SCN $\in \mathbb Z$; CN $=$ SCN \\ 
 \hline
 meron & CN $\in \mathbb Z/2$; SCN $\in \mathbb Z/2$; CN $=$ SCN \\ 
 \hline
  skyrmionium & CN $=0$; SCN $\in \mathbb Z$ \\ 
 \hline
  ``meronium'' & CN $\in \mathbb Z/2$; SCN $\in \mathbb Z/2$; CN $\neq$ SCN \\ 
 \hline
\end{tabular}
\caption{Characterization of different 2D textures by Chern (CN) and staggered-Chern(SCN) numbers. Here, we have introduced the term ``meronium'' to refer to the texture formed out of a meron and a skyrmion with opposite winding numbers.}
\label{tab:chern-nos}
\end{table}

By contrast, the Weyl-planes that separate dipolar and Chern insulting planes, support a novel texture that is a superposition of a skyrmion and a meron with \emph{opposite} chirality.
This leads to a half-integer staggered Chern number, and a \emph{distinct} (in magnitude)  half-integer Chern number.
These features are summarized in Table~\ref{tab:chern-nos}.

\subsection{Flux through Gaussian surfaces}
As a function of $\delta_3$ the oppositely charged Weyl points on the same side of the $k_z$-axis merge to give rise to a Berry dipole at $\delta_3 = 0$ [please see Fig. 1(b) of the main text].
This is a topological quantum critical point. 
Consequently, it would be expected to control the behavior of the system over an  extended energy/momentum-scale  window called a ``critical fan'' (please see Fig.~\ref{fig:crossover}(a)).
The boundary of the critical fan tracks the separation between the Weyl points.
We construct a Gaussian surface centered at one of the Weyl points, and increase its radius to travel along the vertical dashed line in Fig.~\ref{fig:crossover}(a). 
Therefore, at smaller radii we are probing the property of isolated Weyl points and obtain a quantized monopolar Berry flux.
At radii that exceed the separation between the Weyl points, we enter the critical fan, and our probe detects a quantized dipolar flux which characterizes the topological obstruction in the critical fan. 
These behaviors are summarized in Fig.~\ref{fig:crossover}(b).
The existence of this dipolar flux can be attributed to the existence of a Berry dipole   at the topological critical point.
Below we provide the details of the computation.

\begin{figure}
    \centering
\includegraphics[width=12cm]{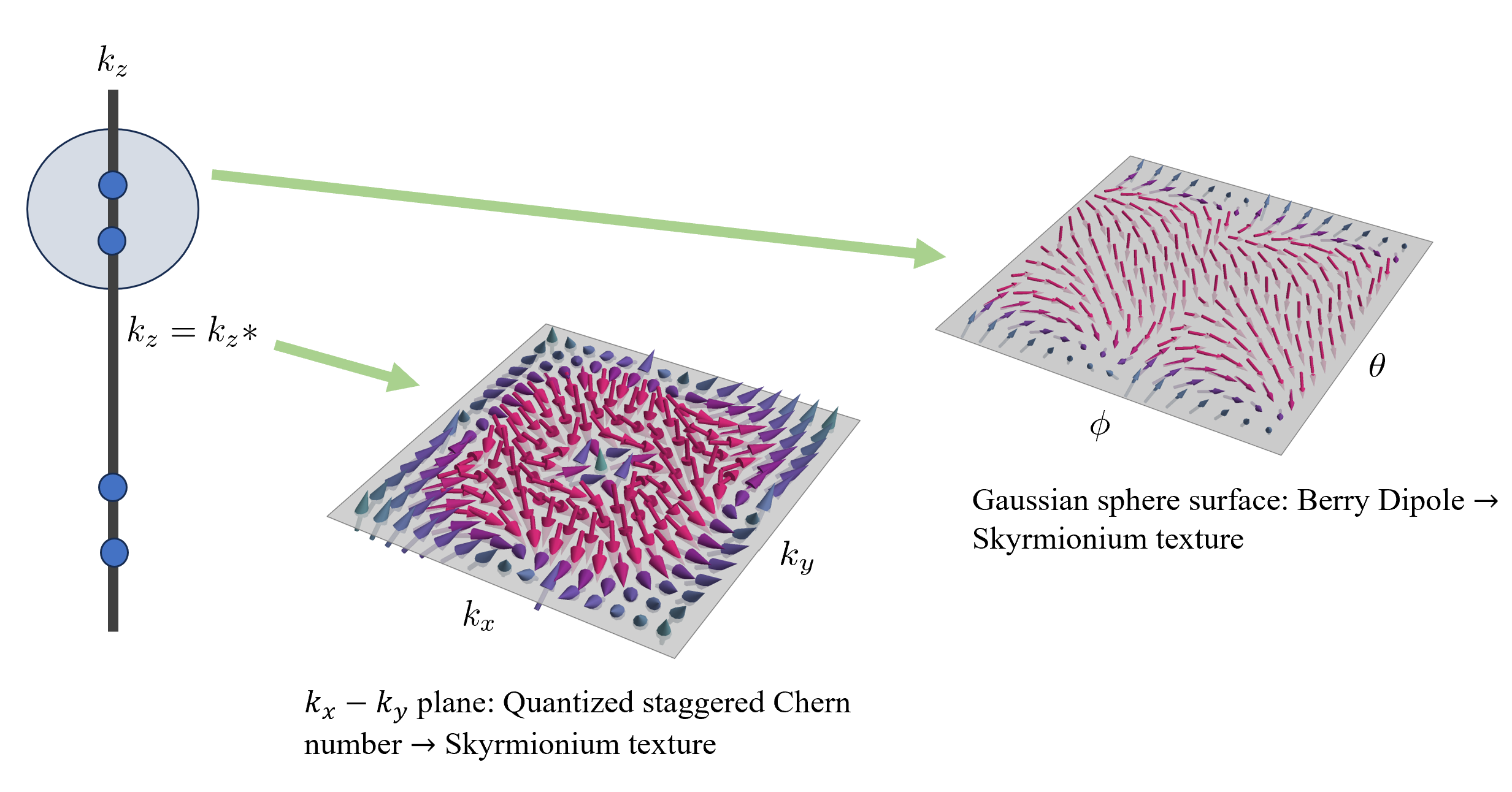}
    \caption{Texture of $|n|$ vector in a generic $k_{x}-k_{y}$ plane supporting quantized dipolar flux ($k_{z}=k_{z}*$ plane) as well as on a gaussian sphere enclosing the Berry dipole. The skyrmionium texture visible in both situations demonstrates the presence of a quantized dipolar flux through a gaussian sphere enclosing the Berry dipole.}
    \label{fig:skyrmionium}
\end{figure}

From the Hamiltonian we determine the Berry curvature
\begin{align}
\bs{\mc B} = \frac{1}{4\pi} (\hat n \cdot \partial_{k_y} \hat n \cross \partial_{k_z} \hat n, 
~~\hat n \cdot \partial_{k_z} \hat n \cross \partial_{k_x} \hat n, 
~~\hat n \cdot \partial_{k_x} \hat n \cross \partial_{k_y} \hat n).
\end{align}
Next, we introduce spherical polar coordinates, $(k_r, \theta, \phi)$ with respect to the Weyl point at $k_z = \cos^{-1}(\Delta - 2 - \dl_3)$, and obtain the radial component of $\bs{\mc B}$,
\begin{align}
\mc B_r(k_r, \theta, \phi) = \bs{\mc B}(k_r \sin{\theta} \cos{\phi}, k_r \sin{\theta} \sin{\phi}, \cos^{-1}(\Delta - 2 - \dl_3) + k_r \cos{\theta}) \cdot (\sin{\theta} \cos{\phi}, \sin{\theta} \sin{\phi}, \cos{\theta}).
\end{align}
The monopolar and dipolar fluxes through the Gaussian sphere centered at the Weyl point are given by, respectively, 
\begin{align}
& \Phi_{\text{mono}}(k_r) = k_r^2 \oint \dd{\theta} \dd{\phi} \sin{\theta} ~\mc B_r(k_r, \theta, \phi) \\
& \Phi_{\text{di}}(k_r) = k_r^2 \oint \dd{\theta} \dd{\phi} \sin{\theta} ~\mc B_r(k_r, \theta, \phi) ~ \text{sign}(\cos{\theta}).
\end{align}
We plot the two fluxes as a function of $k_r$ in Fig.~\ref{fig:crossover}(b).

At $\dl_3 = 0$ the pair of Weyl points on each side of the $k_z$ axis merge to form a quadratic band-crossing point. 
These band singularities carry  Berry-dipolar charge which can be made explicit by ``linearing'' the Hamiltonian close to the band crossing points,
\begin{align}
h_0^{\pm}(\bs q) = \pm \qty[2 t_s t_p \alpha q_x q_z \sigma_1 + 2 t_s t_p \alpha q_y q_z \sigma_2 ] 
+ \qty[t_s^2 \alpha^2 q_z^2 - t_p^2 (q_x^2 + q_y^2)] \sigma_3,
\label{eq:k.p}
\end{align}
where $\bs q$ is the deviation in momentum.
Thus, the dipolar charge can be diagnosed by $\Phi_{\text{di}}$.
{
Our thesis is that a Berry dipole leads to a skyrmionium texture of $\hat n$ on a Gaussian sphere enclosing it, in analogy to a Berry monopole generating a skyrmion texture.
}
In contrast to the Berry dipole that emerges at the topological critical point of Hopf insulators due to an emergent mirror symmetry~\cite{nelson2022}, the Berry dipole obtained here continues to possess a dipolar structure beyond the low-energy  limit in Eq. \eqref{eq:k.p}. 
{
In particular, while an emergent mirror symmetry simplifies the skyrmionium texture, with mirror-related hemispheres of the Gaussian sphere supporting oppositely charged skyrmions, this symmetry is not necessary for a skyrmionium to exist.
The sufficient condition for a skyrmionium texture to exist is the $\hat n$-vector interpolating as follows: north-pole $\to$ south-pole $\to$ north-pole (or vice-versa) along each half-cycle of a compact manifold (eg. 1-torus, 2-sphere, etc).
We demonstrated this explicitly for  generic $k_z$ planes where the skyrmionium texture exists on an 1-torus (i.e. the 2D Brillouin zone).
The skyrmionium texture on the Gaussian sphere can be related to those observed on generic $k_z$-planes by considering a stereographic projection of the sphere to a plane.
In Fig.~\ref{fig:skyrmionium} we portray this connection by plotting a comparison of the texture of $\hat n(\bs k)$ on the Gaussian sphere vs. a $k_z$-plane supporting a quantized Berry dipole flux.
Thus, the Berry dipole in our case is stable against lattice-allowed deformations of $\hat n$, while that obtained at the topological critical point of a Hopf insulator is ``emergent''.
While the existence of the polar-interpolation guarantees the existence of the dipolar flux beyond the low-energy limit, its computation becomes difficult in the absence of an emergent mirror symmetry.
}


\section{1D winding numbers} \label{app:1D-wind}
\begin{figure}
\centering
\subfloat[\label{fig:1D-winding}]{%
  \includegraphics[width=8cm]{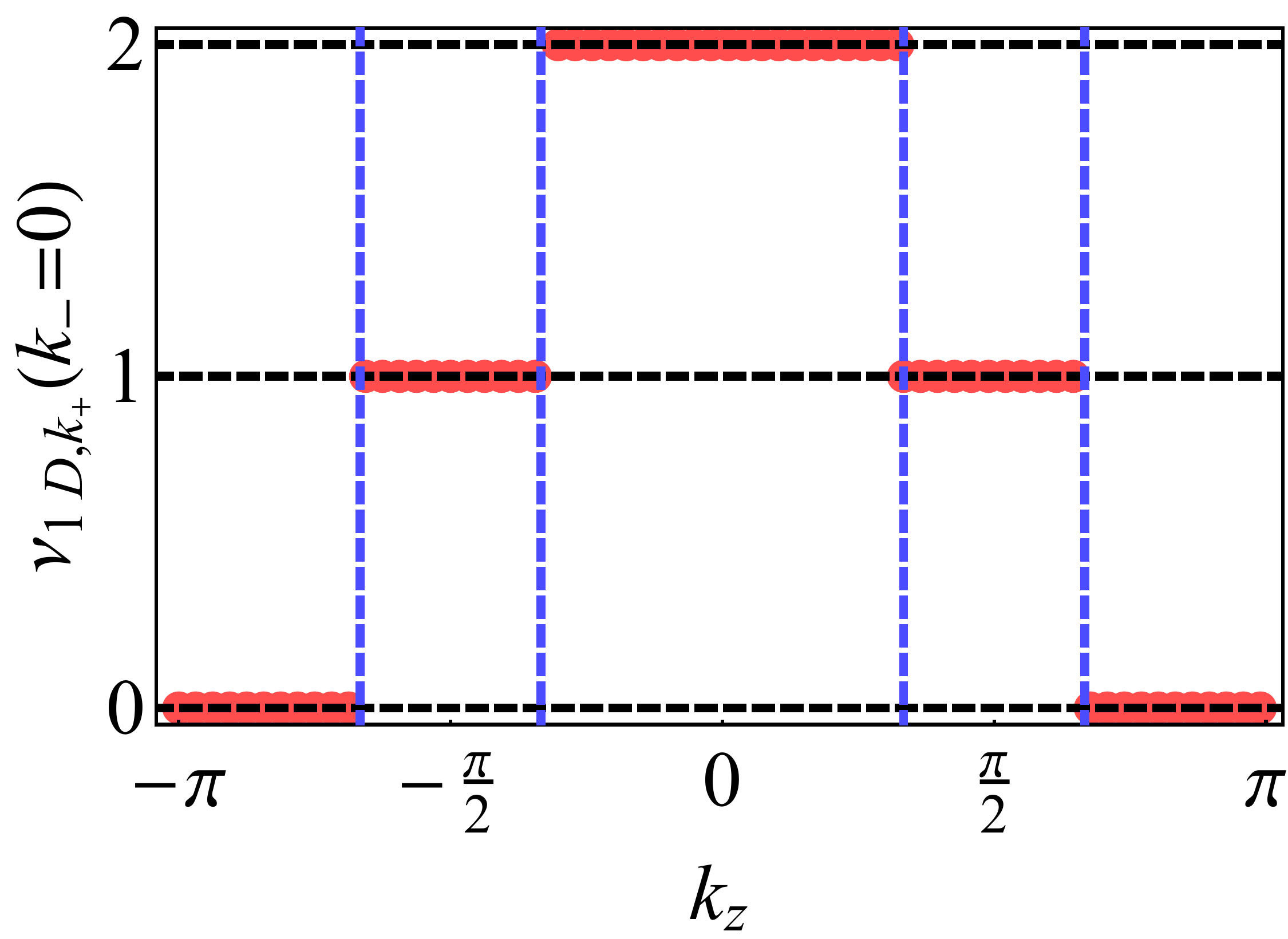}}
\hfill
\subfloat[\label{fig:1D-winding2}]{%
  \includegraphics[width=8cm]{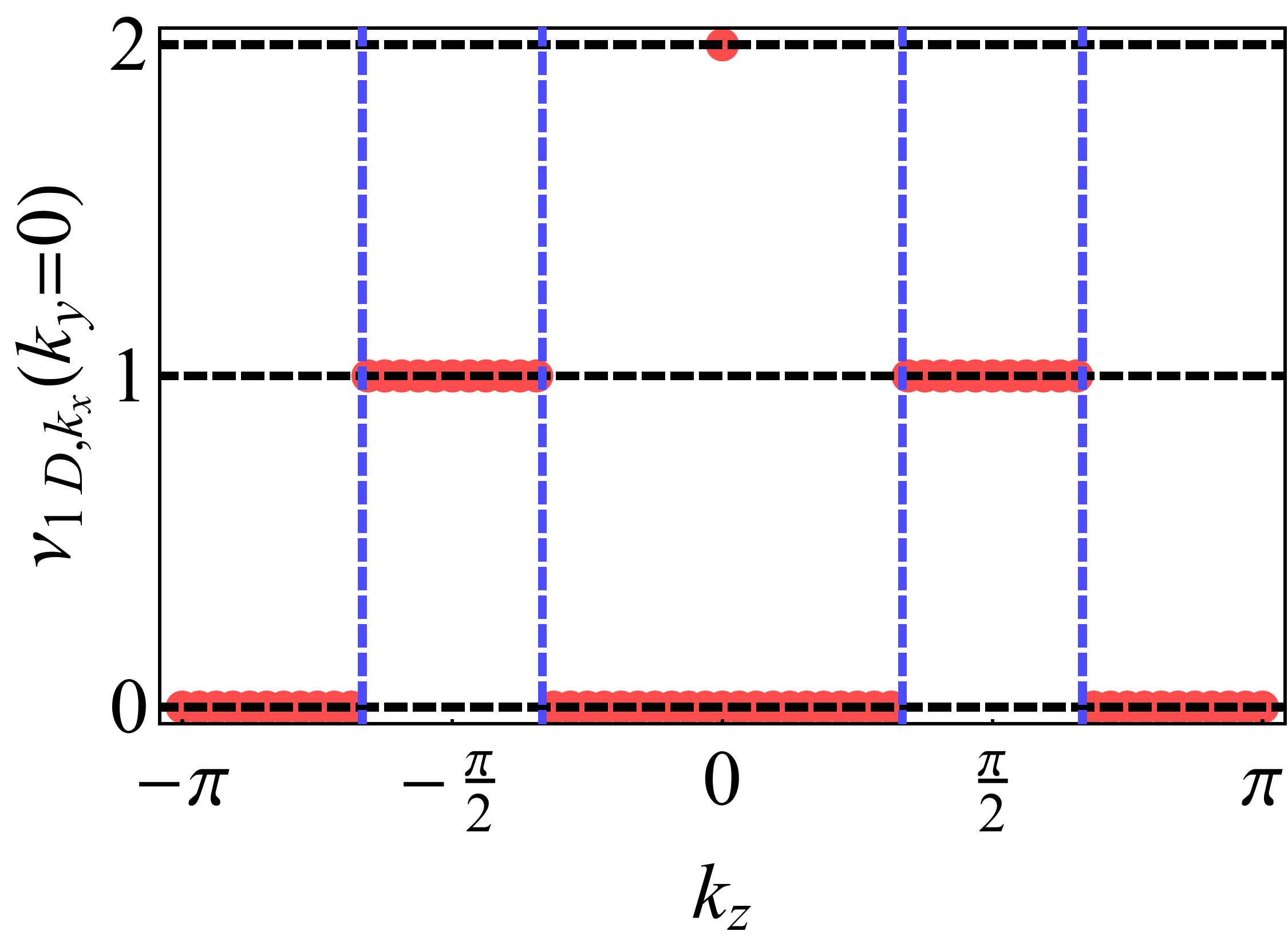}%
}
\caption{One-dimensional winding number along (a) $k_{x}=\pm k_{y}$ and (b) $k_{x}$ axis as a function of $k_{z}$ in dipolar WSM. Location of Weyl points is marked by dashed blue lines. Planes supporting non-trivial Chern number are identified by unit winding number regardless of axis choice. Planes supporting staggered Chern number support doubled winding number along $k_{x}=\pm k_{y}$ axis. Double winding number is recovered along the principal axis in the $k_{z}=0$ plane. This is in correspondence with the existence of a gapless Dirac cone on the (100) and (010) surfaces in this plane.}
\label{fig:winding}
\end{figure}

In order to contrast the topological obstruction between the two types of topologically non-trivial planes, we determine the 1D winding numbers, $\nu_{1D}$, along the $k_x = \pm k_y$ axes of respective planes~\cite{sur2022, yan2019}. The winding number along the $k_{\pm}=k_{x}\pm k_{y}$ axis is computed as\cite{Qi2008}, 
\begin{equation}
    \nu_{1D}=\oint\frac{d k_{\pm}}{2\pi}\sum_{E_{\alpha}(\mathbf{k})<0}(-i)\bra{\psi_{\alpha}}\partial_{k_{\pm}}\ket{\psi_{\alpha}}.
\end{equation}
The result is presented in Fig.~\ref{fig:1D-winding}.
The $k_x = \pm k_y$ axes on the Chern (non-Chern topological insulating) planes support $|\nu_{1D}| = 1$ ($2$).
The $k_z$-planes in the vicinity of $k_z = \pi$ are topologically trivial, since they do not support finite Chern or 1D winding numbers.
The $k_z = 0$ and $\pi$  planes are special, since $u_3$ vanishes on these planes.
Consequently, $k_x$ and $k_y$ axes may be classified by $\nu_{1D}$, in addition to $k_x = \pm k_y$ axes. This is seen in Fig. \eqref{fig:1D-winding2}, and it is this special feature of the plane which protects the gapless Dirac cone seen on the (100) and (010) surfaces in Fig. 3a of the main body. 

\twocolumngrid

\end{document}